# The Discovery of Giant Positive Magnetoresistance in Proximity to Helimagnetic Order in Manganese Phosphide Nanostructured Films

Nivarthana W.Y.A.Y. Mudiyanselage[1], Derick DeTellem[1], Amit Chanda[1], Anh Tuan Duong[2], Tzung-En Hsieh[3,5], Johannes Frisch[3,5], Marcus Bär[3,4,5,6], Richa Pokharel Madhogaria[7], Shirin Mozaffari[7], Hasitha Suriya Arachchige[7], David Mandrus[7], Hariharan Srikanth[1], Sarath Witanachchi[1], and Manh-Huong Phan[1,*]

[1] Department of Physics, University of South Florida, Tampa, Florida 33620, USA

[2] Phenikaa Research and Technology Institute, Phenikaa University, Yen Nghia, Ha-Dong District, Hanoi, 10000, Vietnam

[3] Department of Interface Design, Helmholtz-Zentrum Berlin für Materialien und Energie GmbH (HZB), 12489 Berlin, Germany

[4] Department of Chemistry and Pharmacy, Friedrich-Alexander-Universität Erlangen-Nürnberg (FAU), 91058 Erlangen, Germany

[5] Energy Materials In-situ Laboratory Berlin (EMIL), HZB, 12489 Berlin, Germany

[6] Department X-ray Spectroscopy at Interfaces of Thin Films, Helmholtz Institute Erlangen-Nürnberg for Renewable Energy (HI ERN), 12489 Berlin, Germany

[7] Department of Materials Science and Engineering, University of Tennessee, Knoxville, Tennessee 37996, USA

**The study of magnetoresistance (MR) phenomena has been pivotal in advancing magnetic sensors and spintronic devices. Helimagnets present an intriguing avenue for spintronics research. Theoretical predictions suggest that MR magnitude in the helimagnetic (HM) regime surpasses that in the ferromagnetic (FM) regime by over an order of magnitude. However, in metallic helimagnets like MnP, MR in the HM phase remains modest (< 10%), limiting its application in MR devices. Here, a groundbreaking approach is presented to**

**achieve a giant low-field MR effect in nanostructured MnP by leveraging confinement and strain effects along with spin helicity. Unlike the modest MR observed in bulk MnP single crystals and large-grain polycrystalline films, which exhibit a small negative MR in the FM region (~2%) increasing to ~8% in the HM region across 10–300 K, a grain size-dependent giant positive MR (~90%) is discovered near FM to HM transition temperature ($T_N$ ~110 K), followed by a rapid decline to a negative MR below ~55 K in MnP nanocrystalline films. These findings illuminate a novel strain-mediated spin helicity phenomenon in nanostructured helimagnets, presenting a promising pathway for the development of high-performance MR sensors and spintronic devices through the strategic utilization of confinement and strain effects.**

**Keywords:** Helimagnets; Magnetoresistance; MnP nanostructures; Thin films; Strain effect; Magnetic Sensors; Spintronics

*Corresponding author: phanm@usf.edu

# 1. Introduction

Over the last decade, a great deal of research has been directed towards antiferromagnetic (AFM) spintronics since AFM materials have emerged as promising alternatives to ferromagnetic (FM) materials [1]. In contrast to FM materials, AFM materials possess a zero net magnetic moment, fast dynamics, and robustness against stray fields, positioning them as perspective candidates for applications in information storage technology, high-frequency detection, and energy transport [1–4]. The helical magnetic structure is a form of AFM structure, and the helicity degree of freedom is robust against external disturbances, making it well-suited for information retention [2]. Owing to its helical spin structure, thermomagnetic and thermoelectric functionalities, and superconductivity, manganese phosphide (MnP) is considered a multifunctional material for energy harvesting and spintronics applications [5–10]. It has been reported that bulk MnP undergoes two magnetic phase transitions, from the paramagnetic (PM) to FM state at $T_C$~291 K and from the FM to helimagnetic (HM) state at $T_N$~47 K [11], with the transition influenced by the direction of an external magnetic field [12–14].

In recent years, nanostructuring MnP has emerged as a highly effective strategy for manipulating its magnetic and transport characteristics [6,12,15–20]. Notably, de Andres *et al.* observed significant alternations in low-temperature magnetism, especially in the HM region, in MnP nanocrystals (sized between 15 and 40 nm) in GaP:MnP epilayers and MnP films, as compared to their bulk counterparts (in powder and single crystal forms) [20]. While the $T_C$ remained almost unchanged (~291 K), a large shift in $T_N$ from ~47 K for bulk MnP to ~82 K for the GaP:MnP epilayers was noted. The large increase of $T_N$ up to 100 K was also observed for MnP films epitaxially grown on GaAs substrates [21]. Interestingly, the value of $T_N$ as large as 110 K has recently been reported by Madhogaria *et al.* on highly crystalline MnP nanorod films grown on Si substrates [12]. The enhancement of $T_N$ in MnP/GaAs [21] was attributed to strain induced by lattice mismatch between the epitaxial MnP film and the GaAs substrate. In cases of GaP:MnP epilayers [20] and MnP/Si nanorod films [12], the modified low-temperature magnetic behavior, coupled with the significant $T_N$ upshift, was ascribed to the combined confinement (arising from the nanometric size of nanocrystals/nanorods) and surface effects. It is believed that surface strains induced by tensions in these nanocrystals or nanorods substantially modify the helimagnetic structure, leading to the observed $T_N$ changes, while the bulk FM properties of the nanocrystals or nanorods remain largely akin to those of MnP single crystals, preserving the high $T_C$ values [18,20]. A straightforward model based on localized spins has been proposed to describe the HM structure confinement within MnP nanocrystals [20], which can also be extended to interpret the magnetic behavior of the MnP nanorods [12]. Leveraging their inherent nanoscale attributes and tunable magnetic properties dictated by their high aspect length-to-diameter ratio, highly crystalline MnP nanorod films emerge as particularly promising candidates for a myriad of technological applications [5,6,13].

Magnetoresistance (MR) phenomena in magnetic materials have laid the foundation for the development of magnetic sensors and spintronic devices [22–26]. Current efforts are to engineer magnetic thin films to achieve large low-field MR effects [26]. Exploring MR in helimagnets is of potential interest to AFM spintronics [27]; however, the literature presents limited studies on this topic [13,28–31]. Notably, Zadorozhnyi *et al.* have theoretically predicted a giant MR effect in helimagnets (e.g., $FeGe_2$, $Cu_2OSeO_3$, MnP), when subjected to an external magnetic field perpendicular to the spin spiral propagation axis [31]. Remarkably, the predicted MR magnitude in the HM regime surpasses that in the FM regime by more than an order of magnitude. Experimental validation of this prediction has been achieved in the case of MnP [6,13]. However, owing to MnP's metallic nature, the MR magnitude in the HM phase remains relatively modest (less than 10%), impeding its utilization in MR devices [6,13]. Thus, enhancing the MR effect in helimagnetic systems like MnP is imperative. Previous studies have demonstrated that pressure can alter the magnetic and superconducting properties of MnP [32–34]. However, its impact on the magneto-transport property of nanostructured MnP has not been thoroughly explored.

By harnessing confinement (nanostructuring) and strain effects coupled with spin helicity, we introduce a novel approach to achieve a giant low-field MR effect in helimagnetic systems like MnP. In contrast to the modest MR effect observed in MnP single crystals and large-grain polycrystalline films, characterized by a small *negative* MR effect in the FM region (~2%) that increases to ~8% in the HM region across a temperature range of 10 – 300 K, our study reveals a substantial *positive* MR effect in MnP nanocrystalline films. These films exhibit a giant positive MR effect (~90%) near the FM to HM phase transition temperature ($T_N$ ~110 K), followed by a rapid decline below this temperature, becoming *negative* below ~55 K. The magnitude of the *positive* MR effect varies depending upon the size of MnP grains (below 100 nm). These findings shed light on the novel strain-mediated spin helicity phenomenon in nanostructured helimagnets, offering a promising avenue for designing high-performance MR sensors and spintronic devices by leveraging confinement and strain effects.

2. Results

As shown in **Figure 1a-c**, the XRD analysis reveals an orthorhombic crystal structure with the Pnma space group for MnP-400C, MnP-500C(N), and MnP-500C(S) thin films. The corresponding lattice parameters are $a$ = 5.260 Å, $b$ = 3.174 Å, and $c$ = 5.919 Å for MnP-400C, while they are $a$ = 5.259 Å, $b$ = 3.173 Å, and $c$ = 5.917 Å for MnP-500C(N) and MnP-500C(S). With reference to the lattice parameters of the MnP SC ($a$ = 5.242 Å, $b$ = 3.180 Å, $c$ = 5.903 Å) (**Figure S1**), we estimated 0.34% and 0.27% tensile strain along the *a* and *c* axes, respectively, and 0.19% compressive strain along the *b*-axis, while both MnP-500C samples showed 0.32% and 0.24% tensile strain along the *a* and *c* axes, respectively, and 0.22% compressive strain along the *b*-axis. This observation suggests that the strain effects on all three MnP films, caused by their deposition on Si substrates, are roughly of the same magnitude. The preferred crystallographic orientations of MnP-500C(N) and MnP-500C(S) coincide with each other, while the MnP-400C films show a set of preferred orientations. As displayed in **Figure 1d-f**, XPS measurements confirm the presence of Mn and P in all MnP films investigated, as expected. The O and C signals are attributed to surface oxidation and/or contamination from exposure to ambient conditions prior to XPS analysis. The detailed spectra of Mn $2p_{3/2}$ and P 2p for the three different MnP films are shown in **Figure S2**. The most prominent Mn $2p_{3/2}$ peak was consistently observed at 639.0 eV, and the P $2p_{3/2}$ peak at 129.2 eV, which are associated with Mn-P bonds in MnP [35–37]. The spectral intensity at higher binding energy for Mn $2p_{3/2}$ between 641 and 645 eV and for P 2p between 132 and 136 eV suggest contributions from Mn-O and P-O bonds, respectively [38], in line with the prominent O 1s line seen in the survey spectra (**Figure 1d-f**). To confirm that these contributions were mainly due to surface oxidation, we performed a mild sputter cleaning (SPC) of the MnP samples. Upon SPC, the films exhibited a notable reduction of these high binding energy spectral contributions, as exemplarily evidenced by the Mn $2p_{3/2}$ and P 2p spectra of the SPC MnP-400C sample (see **Figure S3**). Importantly, the main Mn $2p_{3/2}$ and P $2p_{3/2}$ peaks remained at the same binding energy positions of 639.0 eV and 129.2 eV, respectively, compared to the pristine (surface oxidized) sample, indicating an unchanged Mn-P bond like chemical environment.

The SEM images reveal coarse granular structures in MnP-500C(N) and MnP-500C(S), while slightly less granular structure is visible in MnP-400C (refer to **Figure S4a-c**). **Figure 2a-c** shows AFM topography images of MnP-400C, MnP-500C(N), and MnP-500C(S), respectively. The root means square (RMS) surface roughness values for MnP-400C, MnP-500C(N), and MnP-500C(S) are estimated to be ~6.0 nm, ~5.8 nm, and ~5.7 nm, respectively. Noticeably, the AFM images unveil the presence of smallest grains in MnP-400C compared to MnP-500C(N) and MnP-500C(S) films. The average grain sizes are determined to be ~ 39 nm, ~86 nm, and ~121 nm for MnP-400C, MnP-500C(N), MnP-500C(S), respectively. The cross-sectional high-resolution transmission electron microscopy (HRTEM) analysis of MnP-400C reveals distinct crystallite boundaries, with an average crystallite size of approximately 40 nm (see **Figures 2d,c**). **Figure 2f** shows a selected area electron diffraction (SAED) image of MnP-400C, with a set of diffraction points corresponding to the lattice planes of the MnP single crystal. These findings align with the AFM analysis conducted on the same sample, further corroborating the observed results.

**Figure 3a** shows $\rho(T)$ of MnP-400C in the absence of a magnetic field. The linear decrement of $\rho(T)$ from 278 K to 112 K, denoted as regime I, is related to the metallic behavior of MnP. The inset of **Figure 3a** illustrates the temperature-dependent magnetization $M(T)$ under zero-field-cool (ZFC) protocol, indicating that MnP-400C undergoes two magnetic phase transitions from the PM to FM state at $T_C$ ~279 K, and from the FM to HM state at $T_N$ ~110 K. The slope change at ~279 K in $\rho(T)$ further corroborates the PM-FM phase transition within the MnP-400C film. However, the metallic behavior is disturbed by an upturn observed at ~75 K to ~53 K, termed as regime II, followed by a subsequent decrease in resistivity with further temperature reduction, identified as regime III. The broad resistivity minima in $\rho(T)$ for the MnP films are noticeable. Such a broad (metal-insulator-like) transition in the vicinity of the FM-HM transition is not observed for the MnP single crystal (refer to **Figure S5a**). **Figure 3b** displays a similar trend in $\rho(T)$ for MnP-500C(N), with magnetic phase transitions (refer to the inset of **Figure 3b**) occurring at $T_C$ ~300 K (PM-FM) and $T_N$ ~102 K (FM-HM). In contrast, the MnP-500C(S) film, depicted in **Figure 3c**, shows a metallic behavior across the entire temperature range (10-300 K), without the maximum in $\rho(T)$ at $T < T_N$. Such a behavior is similar to that of the MnP SC (refer to **Figure S5a**). This implies that the MnP-500C(S) film possesses bulk-like properties akin to those of the MnP single crystal. As can also be seen in the inset of **Figure 3c**, MnP-500C(S) possesses $T_C$ ~ 304 K and $T_N$ ~74 K. Unlike MnP-400C and MnP-500C(N), the magnetization kept decreasing gradually at temperatures below $T_N$ for MnP-500C(S).

**Figure 3d-f** illustrates the magnetic hysteresis loops $M(H)$ taken at 125, 100, 95, and 50 K for MnP-400C, MnP-500C(N), and MnP-500C(S), respectively, in comparison with the MnP SC (**Figure S5b**). It is worth noticing in **Figure 3d** that for MnP (400 C), the $M(H)$ loops measured $T < T_N$ (~112 K) showed a shrinking behavior at low magnetic fields, suggesting phase coexistence and competition in the HM region. This behavior became more pronounced with lowering temperature, as evidenced by the contrasting $M(H)$ loops obtained at 50 K and 95 K. A similar but less pronounced low field shrinking behavior was also observed at $T < T_N$ (e.g., $T$ = 50 K) for MnP-500C(N). Like MnP-400C, this behavior became more prominent at lower temperatures (at 10 K, see **Figure S6**). In contrast, the $M(H)$ loops for MnP-500C(S) taken at 95 K and 125 K show a similar field dependent magnetic hysteresis behavior, and there appears only a slight deviation in shape of the $M(H)$ loop taken at 50 K. Even at 10 K, the shape of the $M(H)$ loop has not been considerably changed (see **Figure S7**). Referring to **Figure S5c**, the MnP SC shows magnetic phase transitions at $T_C$ ~ 296 K and $T_N$ ~ 52 K, similar to what was reported in the literature [11,39]. It should be noted that the $\rho(T)$ of the MnP SC exhibited a slight, short-range kink at ~ 48-58 K (**Figure S5a**), coinciding with the vicinity of the FM-HM transition. Takase *et al.* observed a sharp drop in $\rho(T)$ at ~ 53 K in MnP whisker samples, which was attributed to the screw-ferro transition,

with the screw phase assigned as the magnetic ground state of the helical spin state of MnP [39,40]. Cheng *et al.* also observed a clear dip in $\rho(T)$ near the FM-HM transition and attributed it to the propagation of a helical magnetic structure[34]. Therefore, the occurrence of a minimum followed by an upturn in $\rho(T)$ observed for MnP-400C and MnP-500C(N) can initially be attributed to the FM-HM transition. The emergence of peaks in $\rho(T)$ around 50 K, just below $T_N$, for these two samples (**Figure 3a,b**) is notable and has not been reported in the literature. Such a peak in $\rho(T)$ is nearly absent in MnP-500C(S) (**Figure 3c**) and MnP SC (**Figure S5a**). The distinct features observed in $\rho(T)$ for MnP-400C and MnP-500C(N) cannot be solely attributed to the FM-HM transition but also to the strain and confinement effects mentioned earlier. This has intricately connected to the competition between magnetic phases at the vicinity, ultimately leading to the attainment of the magnetic ground state in the HM state of MnP.

To further elucidate these intriguing features, we studied the influence of magnetic field on the resistivity of all samples. **Figure 4a** displays $\rho(T)$ of MnP-400C upon the application of magnetic fields up to 7 T. Remarkably, significant variations in resistivity occurred in the vicinity of the FM-HM transition, with substantial suppression, and even disappearance, of the $\rho(T)$ peak at ~53 K under high applied magnetic fields. The deviation of resistivity values at 7 T from the resistivity values at 0 T, denoted as $\Delta\rho = \rho(H=7\text{ T}) - \rho(H=0\text{ T})$, where $\rho(H=7\text{ T})$ and $\rho(H=0\text{ T})$ represent the resistivity at magnetic fields of 7 T and 0 T, respectively, gradually increases with decreasing temperature from 300 K to close to the $T_N$ of MnP-400C, as illustrated in **Figure 4b**. As the temperature further decreases below the $T_N$, $\Delta\rho$ decreases and crosses zero towards negative values, indicating lower scattering centers at 7 T when compared to 0 T.

To gain deeper insights into the $\rho(T, H)$ dependence, in-plane (IP) MR measurements were carried out over a wide temperature range of 10 – 300 K, with the magnetic field ($H$) applied parallel to the plane of the film surface. Here, MR is defined as $MR(\%) = \frac{\rho(H=7\text{ T})-\rho(H=0\text{ T})}{\rho(H=0)} \times 100\%$. The temperature dependence of MR ratio for MnP-400C is displayed in **Figure 5a**. It is worth mentioning that the MR vs. *H* or MR(*H*) curves exhibited a parabolic behavior at $T > T_N$ as the magnetic field was swept from 7 T to -7 T. A low field cusp behavior in MR data started to appear at ~100 K in the lower field range (< 2 T) while it approached a saturation for higher fields (> 2 T). High field saturation behavior of MR started to decrease at lower temperatures (< 75 K) while a second broad peak was maintained until 58 K. This second broad peak was also gradually suppressed at lower temperatures (< 58 K). The crossover from the positive to negative MR with respect to magnetic field can be seen in **Figure S8**. The magnetic field, where the MR became zero, shifted to lower fields at reduced temperatures, and at the lowest measured temperature (e.g., 10 K) the MR became entirely negative regardless of applied magnetic field. This can be reconciled with the phenomenon observed by Suzuki *et al.* on the magnetic field-induced suppression in the additional energy gaps created by the screw spin structure in the conduction band, with applied magnetic fields exceeding the critical field [41]. The *M*(*T*) data of MnP-400C (refer to **Figure S9a**), demonstrates the ability to break down the helical spin structure at 2 T at even at 10 K (refer to **Figure S9b**). The observed increase in MR at 58 K, from 2 to 7 T, in MnP-400C is related to the established FM state. However, the formation of the screw phase and the consequent creation of additional energy gaps in the conduction band has led to higher resistivity under low magnetic fields (< 2 T) when compared to the FM state. Since the energy required to overcome thermal fluctuations at lower temperatures is relatively low, the critical field necessary to stabilize the screw phase has dropped below 2 T. As depicted in **Figure 5b**, a similar field-dependent MR behavior was observed in MnP-500C(N). However, the transition from positive to negative MR occurred at a higher temperature (~75 K) compared to MnP-400C (~55 K). The onset of negative

MR in both MnP-400C (~55 K) and MnP-500C(N) (~75 K) aligns with the low-temperature resistivity peaks (~54 K and ~74 K) observed for these two samples (**Figure 3a,b**). This alignment suggests that the temperature range where the broad resistivity peaks were observed in the $\rho(T)$ of MnP-400C and MnP-500C(N) corresponds to the temperature range where the sign of MR switches from positive to negative. It is worth mentioning that over the temperature range of 10 - 300 K, a negative MR was consistently observed for MnP-500C(S) without any evidence of a sign change, as illustrated in **Figure 5c**. The negative MR behavior was also observed in our MnP SC (refer to **Figure S10a**), which agrees with the existing literature for bulk MnP [41].

This striking feature can be better highlighted in **Figure 6** for all three samples. **Figure 6a** depicts the temperature-dependent MR of MnP-400C at a magnetic field of 7 T. At 300 K, the MnP thin film exhibited an MR effect of approximately ~15%, which increased as temperature decreased, reaching ~90% at 110 K (~$T_N$), before declining to around 0% at 58 K. A negative MR effect was observed at lower temperatures (<58 K), mirroring the trend observed in $\Delta\rho$ obtained from $\rho(T)$ measurements. Similarly, **Figure 6b** shows the temperature-dependent MR of MnP-500C(N), following a similar trend to MnP-400C but with lower magnitudes of MR (~40% at 125 K). Conversely, **Figure 6c** illustrates that MnP-500C(S) exhibited negative MR at 7 T across the temperature range, albeit with lower magnitudes compared to the other two films. Notably, the highest MR value (~90% at ~110 K) is observed in MnP-400C in the vicinity of the FM-HM transition, contrasting with MnP-500C(N) (MR ~ 40%). These MR values are much greater compared to those obtained for MnP-500C(S) and MnP SC samples.

To explain the negative magnetoresistance originating from the spin-dependent scattering of carriers by localized magnetic moments, Khosla and Fischer used a semi empirical formula, which is given by [42],

$$MR\ (H) = -a^2\ \ln(1 + b^2 H^2) \qquad (1)$$

where, $a = [A_1 J_{ex} D(E_F)\{S(S+1) + \langle M^2 \rangle\}]^{1/2}$, $b = \left[1 + 4\pi^2 S^2 \left(\frac{2 J_{ex} D(E_F)}{g}\right)\right]^{\frac{1}{2}} \left(\frac{g \mu_B}{\beta k_B T}\right)$, $A_1$ represents a constant for the spin dependent scattering contribution to MR(*H*), $J_{ex}$ is the s-d exchange integral. $D(E_F)$ is the density of state at fermi level, *S* represents the spin of the localized magnetic moment, $\langle M^2 \rangle$ is the average of the squared magnetization, *g* is the Lande g-factor, $k_B$ is the Boltzmann constant, $\mu_B$ is the Bohr magneton and T is the temperature. We used this semi empirical formula to fit all the negative MR(*H*) data of the MnP SC and MnP-500C(S) samples. As evidenced by the fits, Eq. (1) does not fully capture the negative MR behavior of MnP SC and MnP-500C(S). To improve the fitting, we employed the modified Khosla and Fischer (KF) model as given in Eq. (2) [43]. This approach accounts for the positive magnetoresistance MR(*H*) observed in the MnP SC and MnP-500C(S) samples, which reflects contributions from two-carrier conduction channels where the negative MR(*H*) is dominant.

$$MR\ (H) = -a^2\ \ln(1 + b^2 H^2) + \left(\frac{c^2 H^2}{1 + d^2 H^2}\right) \qquad (2)$$

The coefficients $c^2$ and $d^2$ are given by, $c^2 = \frac{\sigma_1 \sigma_2 (\mu_1 + \mu_2)^2}{(\sigma_1 + \sigma_2)^2}$ and $d^2 = \frac{(\sigma_1 \mu_2 - \sigma_2 \mu_1)^2}{(\sigma_1 + \sigma_2)^2}$, where $\sigma_i$ and $\mu_i$ represent the conductivity and mobility of i[th] carrier channel, respectively. The MR(*H*) data were well-fitted using Eq. (2), with the fitting results for MnP SC and MnP-500C(S) shown in **Figure 7a,b** and **Figure S10a,d**, respectively. The temperature dependence of the fitting parameters is shown in **Figure S10b,c** for MnP SC and in **Figure S10e,f** for MnP-500C(S). Evidently, the fitting parameters show that |a| > |c| at higher temperatures (>75 K) for both MnP SC and MnP-500C(S). This indicates that the negative MR(*H*) is primarily governed by the spin-dependent scattering, which significantly influences the electronic transport in this temperature region. On the other hand, at lower temperatures (< 75 K), |a| < |c|, indicating a significant contribution from the two-carrier conduction channels, which results in a positive MR(*H*). This contribution may stem from the surface oxidation of the MnP samples, as indicated by the EDS data for MnP SC (**Figure S11**) and XPS data for MnP-500C(S) (**Figure 1f**). If the surface oxidation observed in XPS were responsible for the sign change and higher MR(*H*) values near the FM-HM transition, the MR(*H*) behavior of SC and MnP-500C(S) should be similar to that of MnP-500C(N) and MnP-400C, since all three films exhibit a similar surface oxidation as indicated by the XPS spectra. To confirm the pattern of |a| and |c| fitting parameters in MnP-500C(N) and MnP-400C, we performed fittings at a low temperature (<75 K), which showed a crossover from positive to negative MR(*H*) as shown in **Figure 7c,d**. Notably, the fitting yielded parameters |a| < |c| in this temperature range, consistent with the trend observed in MnP SC and MnP-500C(S). Therefore, the observed higher MR, ~40 % (MnP-500C(N)) and ~ 90% (MnP-400C) has negligible contribution from the two-carrier conduction channel, excluding any (major) impact of the observed surface oxidation.

Given that MnP-400C showcases the most intriguing magnetic, electrical, and magneto-transport properties, we fit its $\rho(T)$ data to theoretical models to further understand the temperature dependent scattering mechanisms within the nanostructured MnP film. We recall that the variation of resistivity with temperature captures the temperature dependence of scattering mechanisms that dominate at different temperature regimes. To understand the scattering mechanism, we have analyzed the temperature dependent resistivity in regime (I), (II), and (III) of MnP-400C, as illustrated in **Figure 3a**. In a ferromagnet, resistivity results from the scattering of conduction electrons from defects or impurities, phonons, and magnons. According to Matthiessen's rule, the total resistivity ($\rho$) can be expressed as the sum of contributions from all these scattering processes [44]:

$$\rho(T) = \rho_0 + \rho_P(T) + \rho_e(T) + \rho_M(B,T), \qquad (3)$$

where $\rho_0$ is the temperature-independent scattering resistivity due to domain, grain boundaries, defects, or impurities[44,45]. $\rho_P$ is described by the Bloch-Gruneisen formula for contribution of acoustic phonons[46]. $\rho_P$ can be $T$ or $T^5$ dependent for $T \gg \theta_D$ or $T \ll \theta_D$, respectively. Here $\theta_D$ is the Debye temperature[44] and it is greater than ~ 750 K for MnP[47]. $\rho_e(T)$ is attributed to inelastic and elastic electron-electron scattering while $\rho_M(B,T)$ to the electron-magnon scattering [44].

The transport mechanism responsible for the metallic ferromagnetic region in MnP-400C, regime I and III, has been elucidated through fitting the resistivity data using

$$\rho(T) = \rho_0 + \rho_{e-Ph}T^5 + \rho_{e-e}^{el}T^{0.5} + \rho_{e-e/M}T^2 \qquad (4)$$

Since the electron-magnon and inelastic electron-electron scattering contributions exhibit a $T^2$ dependence, their related effect is represented by a single term, $\rho_{e-e/M}$, where this term captures the contributions of both mechanisms. Here, $\rho_{e-Ph}$ and $\rho_{e-e}^{el}$ indicate the temperature-dependent resistivity coefficients for electron-phonon inelastic scattering and electron-electron elastic scattering mechanisms, respectively. Here the electron-electron elastic scattering term accounts for the Coulombic interaction between the charge carriers [48]. The quality of the fitting was assessed based on the values of $r^2$, the chi square of the fitting, and the parameter dependencies. **Figure 8a** shows the experimental $\rho(T)$ data and the fitted curves using Eq. (4) for the temperature range of 125-275 K. The fits were made for $\rho(T)$ under 3 and 7 T (refer to **Figure S12a,b**).

From the fitting curves, we extracted the values of $\rho_0$, $\rho_{e-P}$, $\rho_{e-e}$, and $\rho_{e-e/M}$, with **Table 1** listing these parameters under 0, 2, 3, and 7 T. These findings illustrate that, $\rho_0$ and $\rho_{e-e}$ underscoring the significant roles played by grain boundaries and elastic electron-electron scattering processes in the temperature-dependent conduction within the metallic ferromagnetic region of the film across all magnetic fields[49]. Notably, $\rho_0$ increases with an increasing magnetic field, indicating that the magnetic field enhances the resistivity stemming from grain boundaries, which corroborates the observations in the $\rho(T)$ data under magnetic fields, as depicted in **Figure 4**.

The temperature-dependent resistivity fitted in regime (III) also aligns well with Eq. (4), **Figure 8b**. Analysis of the fitting parameters under zero magnetic field indicates that $\rho_0$, and $\rho_{e-e}^{el}$ highlighting the significant roles played by grain boundary and electron-electron elastic scattering within temperature regime III. **Table 2** displays the values of parameters obtained for magnetic fields of 0, 2, 3, and 7 T within the temperature range of 10 - 44 K, with the low-temperature fitting included in **Figure S13a,b**. Within this temperature regime, $\rho_0$ decreases with an increasing magnetic field, indicating that the magnetic field suppresses the resistivity arising from grain boundaries within this temperature region. This observation is further supported by the negative $\Delta\rho$ (**Figure 4b**) and negative MR (**Figure 6a**). Scattering is more pronounced at low temperatures, ranging from 10 to 55 K, when compared to the temperature region of 125-275 K. The additional energy gaps created in the conduction band in the HM phase has resulted in an increase in resistivity when compared to that of the FM state [50].

To understand the resistivity minimum followed by an upturn in regime II of MnP-400C film, the data was analyzed based on three different mechanisms; (1) quantum interference effect (QIE) such as weak localization (WL) [51], (2) localization caused by electron-electron elastic interaction (EEI) [48,52–54], (3) Kondo like scattering based on spin-dependent scattering [55], and (4) combination of Kondo and EEI effects. In the WL case, the temperature-dependent resistivity can be expressed by the equation (5).

$$\rho(T) = \rho_0 + \rho_{e-Ph}T^5 + \rho_{e-e}^{el}T^{0.5} + \rho_{WL}T^{-0.5} \tag{5}$$

$$\rho(T) = \rho_0 + \rho_{e-Ph}T^5 + \rho_{e-e/M}T^2 + \rho_K \ln(T) \tag{6}$$

$$\rho(T) = \rho_0 + \rho_{e-Ph}T^5 + \rho_{e-e}^{el}T^{0.5} + \rho_K \ln(T) \tag{7}$$

Equations (6) and (7) represent the fitting models considering Kondo-like scattering and the combination of Kondo like scattering and EEI, respectively. Here, $\rho_{WL}$ and $\rho_K$ are the coefficients of WL and Kondo terms, respectively. The deviation of the fitting results from the temperature-dependent resistivity data within region II (**Figure S14**), when accounting for the WL term using Eq. (5), confirms the minimal contribution of the WL effect in the MnP-400C film. Fitting attempts for region II were performed considering various models: EEI (refer **Figure S15a**), Kondo (refer **Figure S15b**), a combination of EEI and Kondo (refer **Figure 8c**), along with other inelastic scattering mechanisms described in Eq. (2). All three fitting models produced virtually indistinguishable results. The statistical parameters associated with the fits - specially the reduced $\chi^2$ and $r^2$ values – are very similar, as shown in **Table S2**. However, the combination of EEI and Kondo scattering in the fitting results in a lower residual resistivity within the temperature range of 70-125 K, as illustrated in **Figure 8d**. In this temperature range and near to the resistivity upturn, Kondo-like scattering and electron-electron elastic scattering competes. Additionally, the magnitude of $\rho_K$ is an order of magnitude greater than $\rho_{e-e}^{el}$ when using Eq. (7), indicating that Kondo-like scattering dominates over electron-electron elastic scattering in this region [56].

3. Discussion

Consistent with previous findings by de Andres *et al.* [20], we have observed that polycrystalline MnP films with nanometric grains (below 100 nm) exhibit enhanced strain and disorder at the grain boundaries, affecting the helimagnetic order, while the interior of the grains remains relatively unaffected compared to bulk MnP single crystals. This explains the observed large increase of $T_N$ and the preservation of $T_C$ in the nanocrystalline MnP film (refer to the inset of **Figure 3a** and **Figure S5c**). The average size of grains in MnP-400C ($d \sim 39$ nm) is smaller compared to MnP-500C(N) ($d \sim 86$ nm), suggesting enhanced strain and more disorder at the grain boundaries in the former. This accounts for the higher value of $T_N$ obtained for MnP-400C. The increased disorder in MnP-400C could also lead to the formation of more scattering centers, coupled with the strain and confinement effects, contributing to the escalation of the MR ratio. The emergence of a new metal-insulator-like transition peak around 50 K in the $\rho(T)$ curve (see **Figure 3a, b**) can be attributed to the combined strain and confinement effects that are significant in MnP-400C and MnP-500C(N). A similar phenomenon has been noted in patterned thin films of $La_{1-x}Pr_xCa_{0.375}MnO_3$ [57]. When the grain size exceeds a critical size (> ~100 nm), however, the strain and confinement effects become less significant. Consequently, large-grain polycrystalline films such as MnP-500C(S) ($d \sim 121$ nm) exhibit a bulk-like $\rho(T)$ or MR behavior similar to that observed for MnP SC. Indeed, in the $\rho(T)$ curves for MnP-500C(S) (**Figure 3c**) and MnP SC (**Figure S5a**), no such metal-insulator-like transition peak around 50 K is observed. Similar to MnP SC, the *negative* MR behavior is observed across all temperatures from 10 to 300 K. On the other hand, the presence of the confinement and strain effects within the FM regime in MnP-400C and MnP-500C(N), governing the transport properties of the films, is what led to the *positive* MR behavior in this FM region, unlike MnP SC and MnP-500C(S). In polycrystalline manganite films, the enhancement in MR has been attributed to spin-dependent scattering at grain boundaries [58]. A similar perspective is anticipated in our nanocrystalline MnP thin films. Our findings shed light on the novel strain-mediated spin helicity phenomenon in nanostructured helimagnets, paving the way for the development of high-performance MR sensors and spintronic devices by strategically harnessing confinement and strain effects. Theoretical prediction by Zadorozhnyi *et al.* [31] aligns closely with our experimental observations for MnP SC and MnP-500C(S), revealing that the magnitude of MR in the HM regime exceeds that in the FM regime by nearly an order of magnitude. Our discovery of the giant *positive* MR effect in MnP-400C, proximal to the FM-HM transition, underscores the significance of incorporating strain and confinement effects into the theoretical model to elucidate the MR phenomenon in nanostructured helimagnets.

## 4. Conclusions

Our study sheds light on the potential of MnP nanostructured thin films as promising candidates for antiferromagnetic spintronics. By leveraging the rich magnetic properties and helical spin structure inherent in MnP, we have demonstrated the tunability of magnetic properties through the utilization of nanostructured MnP. Our investigation highlights the crucial role of crystallite sizes in modulating the magnetic behavior, temperature dependent resistivity and MR effects of MnP thin films. The observed resistivity minima and pronounced high positive MR effects in MnP nanostructured thin films underscore the significance of confinement and strain effects, magnetic phase coexistence and competition in governing the electron transport properties. By elucidating these relationships, we contribute to advancing the fundamental understanding of magnetic nanomaterials and their potential applications in advanced electronic and spintronic devices. Moving forward, further studies are warranted to explore additional parameters influencing the magnetic and transport properties of nanostructured MnP thin films, such as doping, defect engineering, and interface effects. Additionally, the integration of nanostructured MnP thin films into functional devices for spintronic and magnetic memory applications represents an exciting avenue for future research. In summary, our work represents a significant step towards harnessing the full potential of nanostructured helimagnets like MnP thin films in antiferromagnetic spintronics, offering new insights into their tunable magnetic properties and paving the way for the development of innovative spintronic devices.

**Experimental Methods**

MnP thin films were grown on Si (1 0 0) substrates using molecule beam epitaxy under different growth conditions, while maintaining consistent thickness for temperature and magnetic field dependent resistivity studies. The details of sample growth are discussed elsewhere[6]. Here forth, films grown at a substrate temperature of 400 $^0$C are denoted as MnP-400C, while two distinct series of MnP thin films grown at 500 $^0$C substrate temperature are labeled as MnP-500C(N) and MnP-500C(S), respectively. MnP single crystals (SC), which were grown by the Sn flux method, were also employed to investigate the transport property variations from bulk to low dimensions of MnP under identical measurement conditions and procedures.

The crystallinity, crystal structure, and space group of MnP films were investigated using X-ray diffractometer (XRD- RIGAKU SmartLab) with Cu Kα source at room temperature. High resolution transmission electron microscopy (HRTEM-Tecnai TF20) was used for cross sectional imaging of MnP-400C film to conclude the crystallite size. Morphology, topography, and surface roughness analyses of the MnP films were conducted using scanning electron microscopy (SEM-JEOL JSM-6390LV) and atomic force microscopy (AFM-Hitachi AFM5300E) at room temperature under vacuum. The average grain size analysis was performed using Gwyddion software. The presence of elements was further verified by X-ray photoelectron spectroscopy (XPS) using a monochromatized Al K$_\alpha$ source from SPECS (XR 50 M X-ray source with a FOCUS 600 X-ray monochromator) providing an excitation energy of 1486.58 eV. The kinetic energy of photoemitted electrons was determined using a hemispherical electron analyzer from SPECS (PHOIBOS 150) equipped with a channeltron detector (MCD-9). After initial XPS characterization, for the studied samples the surface oxide / contamination layer was removed by a mild sputter cleaning (SPC) conducted by a SPECS IQE 11/35 ion source oriented perpendicular to the sample surface using 200 eV $Ar^+$ ions and an emission current of 7.29 mA for 30 minutes.

The magnetic properties of the above-mentioned samples were investigated by vibrating sample magnetometry (VSM) option in physical property measurement system (PPMS) in Dynacool. Temperature dependent resistivity ($\rho(T)$) and magnetoresistance (MR) were measured using a four-probe configuration and the DC resistivity option in the PPMS. The applied magnetic field was oriented in a plane parallel (IP) and perpendicular (OP) to the plane of the MnP thin films, with the current flowing perpendicular to the plane of the film with respect to the applied magnetic field. For MnP SCs, a four-probe configuration was utilized, with the current set-in *c*-axis direction and the magnetic field in the *ab*-plane. $\rho(T)$ was obtained for magnetic fields of 0 T, 2 T, 3 T and 7 T. For MR analysis, the magnetic field was swept from +7 T to -7 T across a temperature range of 10 K to 300 K. All structural characterization, magnetic, and transport data related to the MnP SC are available in the supplementary information (SI).

## Acknowledgements

Research at USF was supported by the U.S. Department of Energy, Office of Basic Energy Sciences, Division of Materials Sciences and Engineering under Award No. DE-FG02-07ER46438. N.W.Y.A.Y. Mudiyanselage acknowledges the German Academic Exchange Service (DAAD) for granting a RISE professional scholarship that allowed her to perform XPS measurements in the Energy Materials In-Situ Laboratory Berlin (EMIL) at the Helmholtz-Zentrum Berlin für Materialien und Energie GmbH. D. M., H. W. S. A., R. P. M., and S. M. acknowledge the support from the Gordon and Betty Moore Foundation's EPiQS Initiative, Grant GBMF9069. H.S. thanks the Alexander von Humboldt Foundation for a research award.

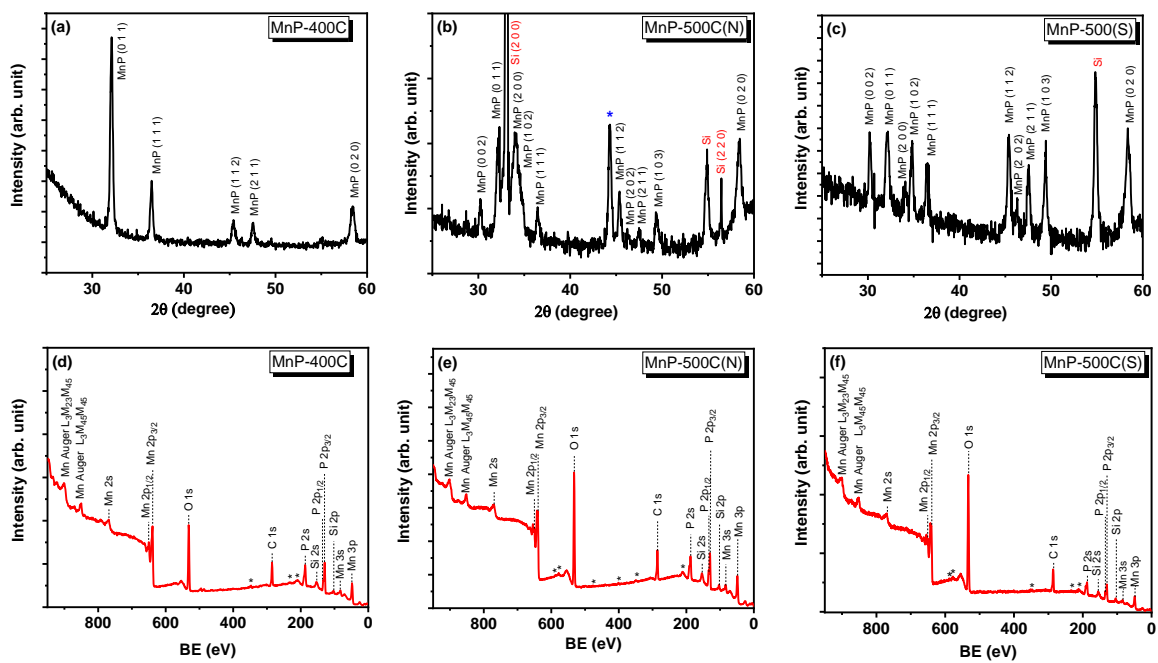

**Figure 1**. X-ray diffraction (XRD) patterns of (a) MnP-400C, (b) MnP-500C(N), and (c) MnP-500C(S); X-ray photoelectron (XPS) survey spectra of (d) MnP-400C, (e) MnP-500C(N), and (f) MnP-500C(S) recorded using a monochromatized Al K$_\alpha$ source (hv=1486.58 eV). The asterisk (blue) in (b) indicates the XRD signal from the sample holder.

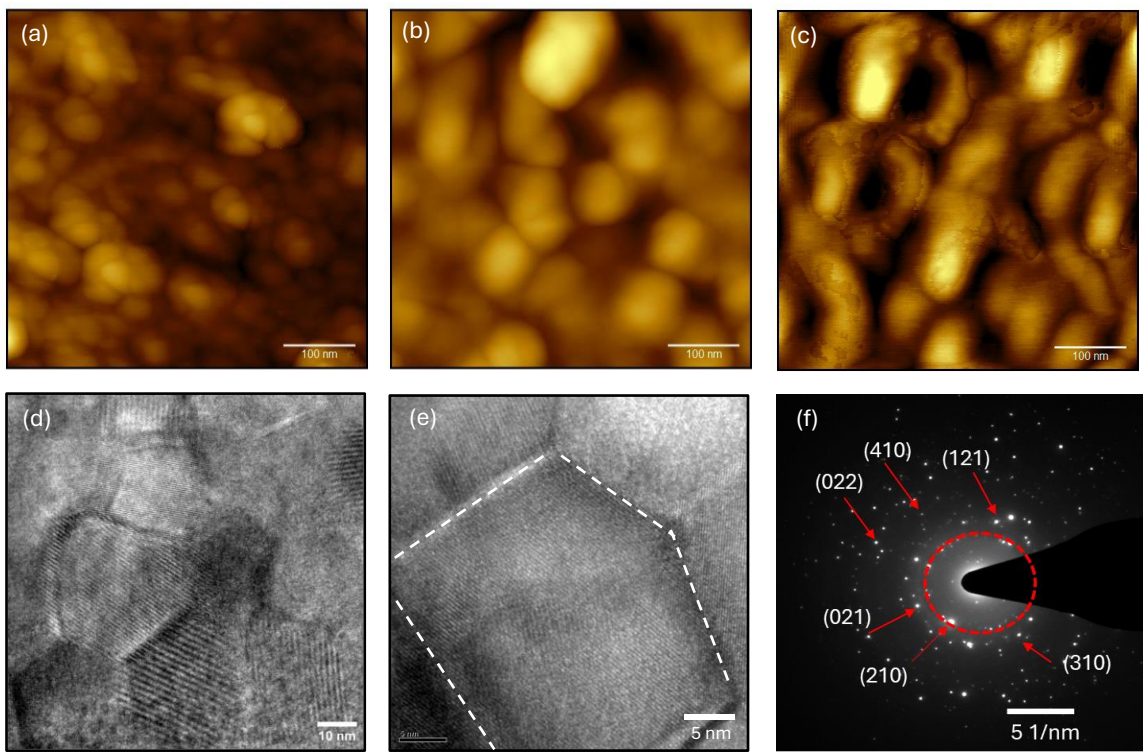

**Figure 2**. Surface morphology by AFM for (a) MnP-400C, (b) MnP-500C(N), and (c) MnP-500C(S); (d, c) Cross-sectional HRTEM images of MnP-400C with different modifications, with dot lines showing grain boundaries; (f) the SAED image of the MnP-400C film.

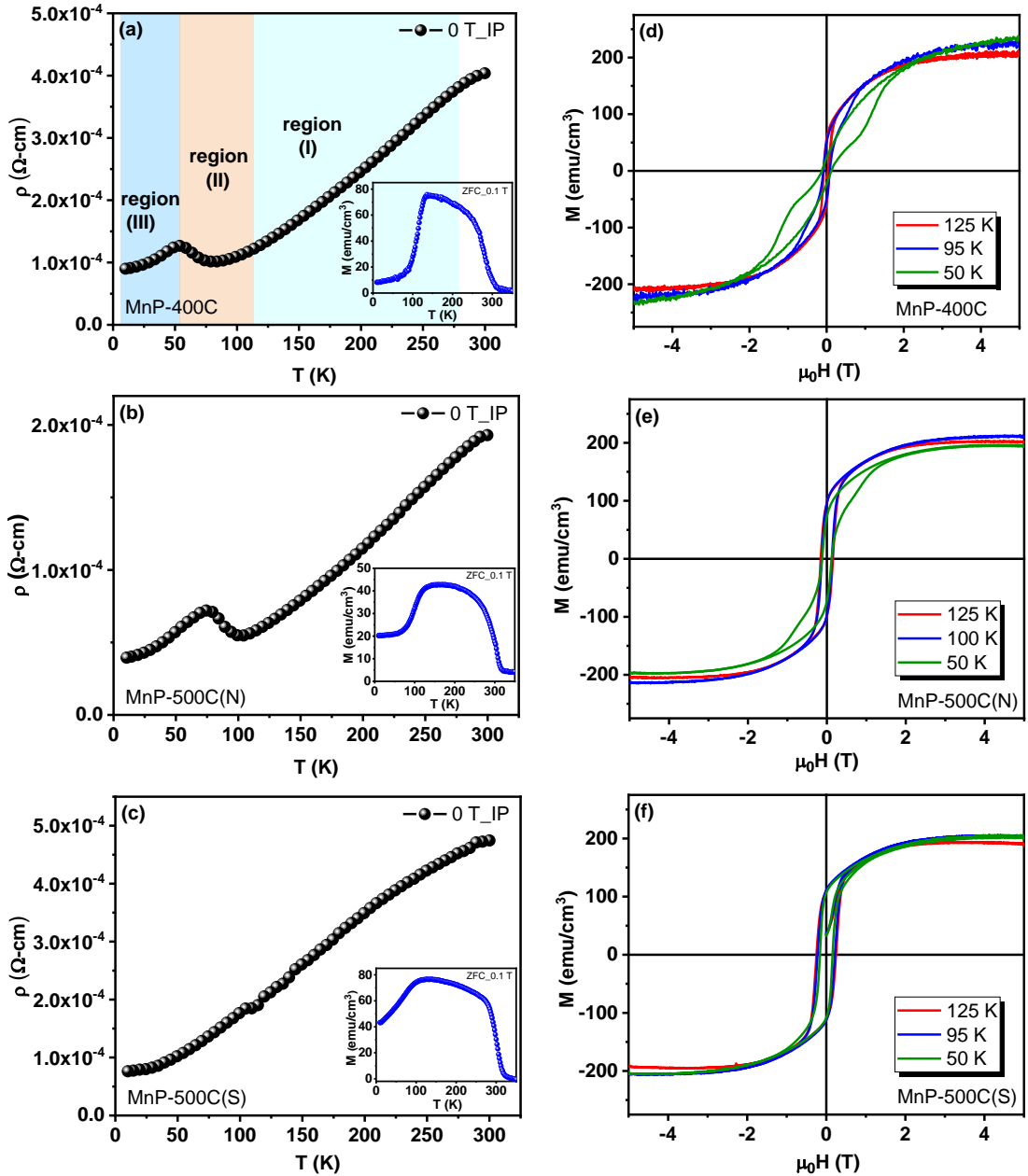

**Figure 3**: The temperature dependent resistivity under zero field for a) MnP-400C, b) MnP-500C(N), and c) MnP-500C(S); Isothermal magnetic hysteresis *M*(*H*) loops of d) MnP-400C, e) MnP-500C(N), and f) MnP-500C(S). Insets of Figure 3a-c show the temperature dependent magnetization M(T) curves under field-cooled measurement protocol for the corresponding samples.

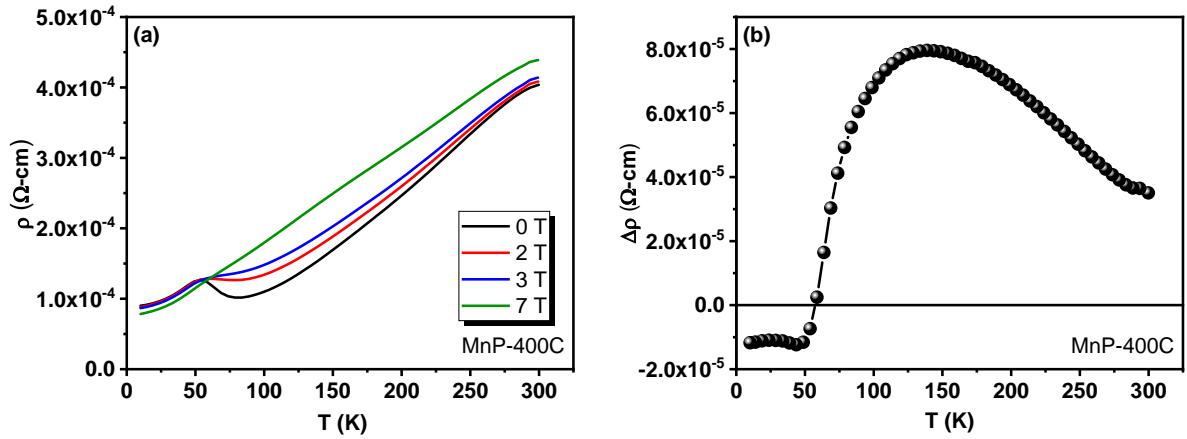

**Figure 4**. (a) Temperature dependent resistivity of MnP-400C under magnetic fields of 0, 2, 3 and 7 T; (b) Relative change in resistivity under 7 T compared to 0 T as a function of temperature, $\Delta\rho = \rho\,(H = 7\,T) - \rho\,(H = 0\,T)$.

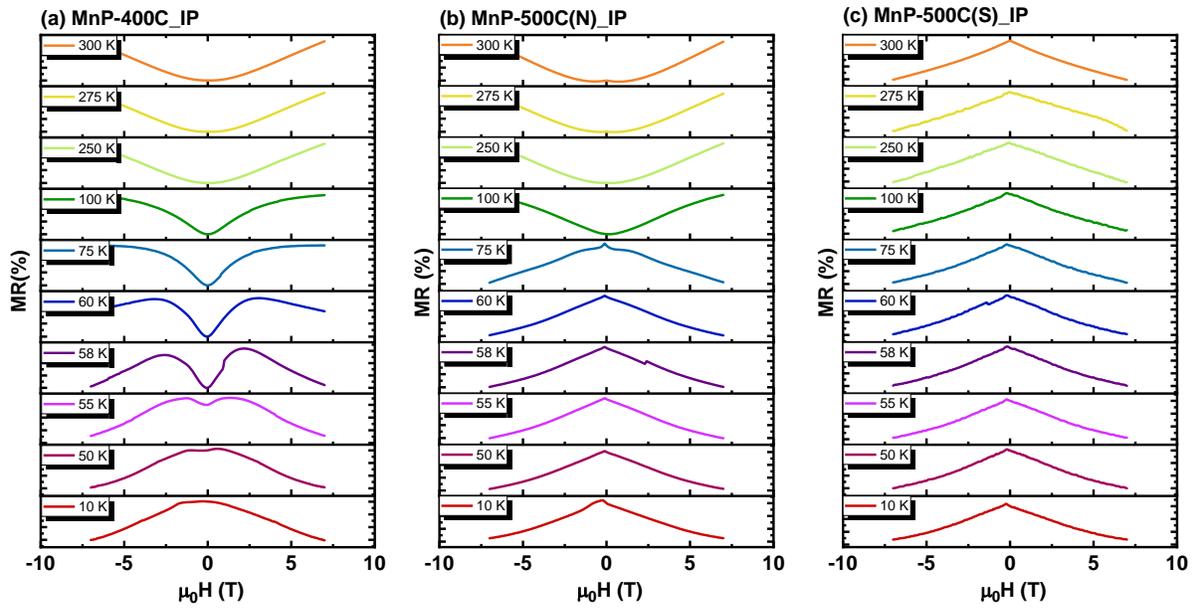

**Figure 5**. Magnetic field dependent magnetoresistance for (a) MnP-400C, (b) MnP-500C(N), and (c) MnP-500C(S).

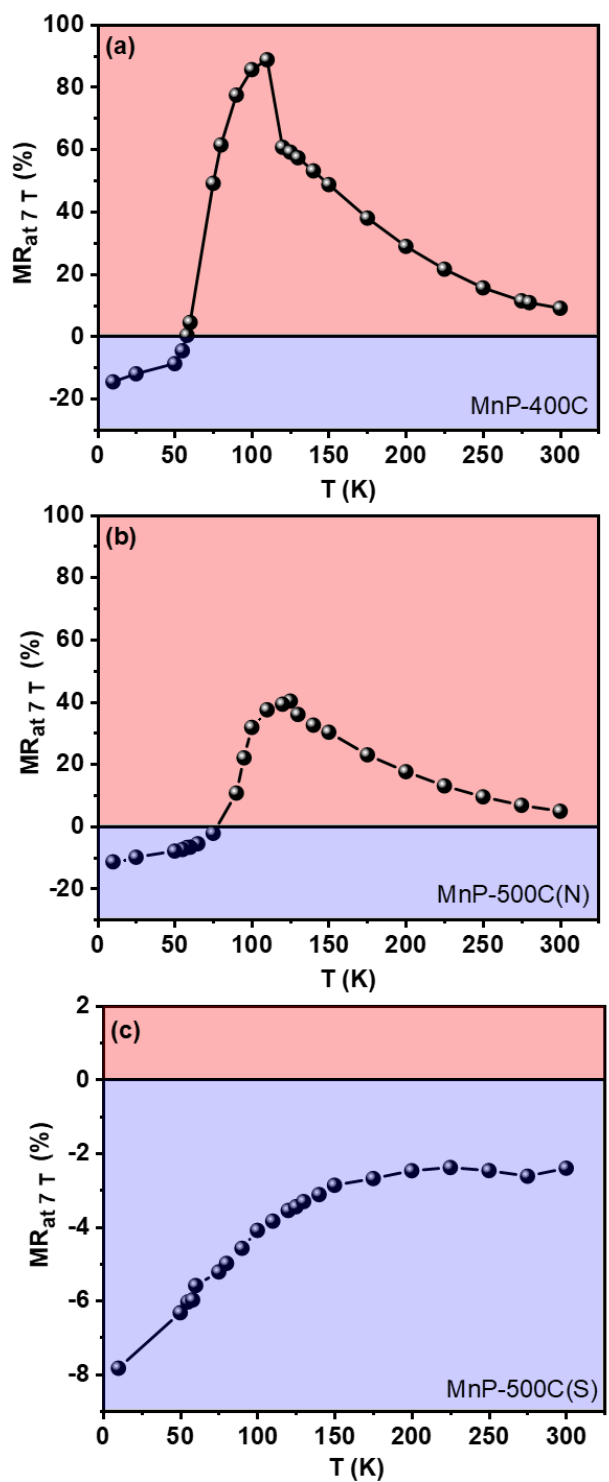

**Figure 6**. Temperature dependent magnetoresistance ratio at 7 T for (a) MnP-400C, (b) MnP-500C(N), and (c) MnP-500C(S).

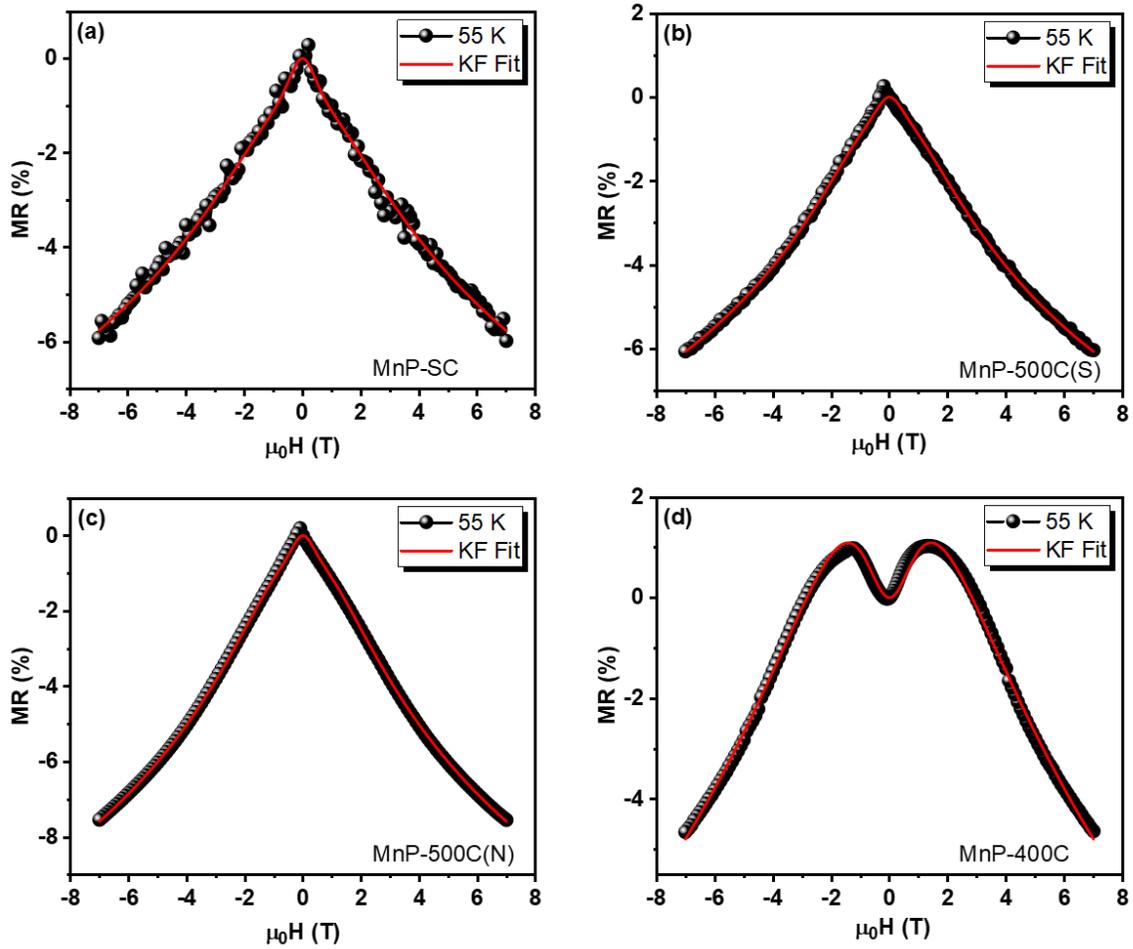

**Figure 7**. Magnetic field-dependent magnetoresistance at 55 K of (a) MnP-SC, (b) MnP-500C(S), (c) MnP-500C(N) and (d) MnP-400C. The solid (red) lines represent the fitting curves using Equation (2).

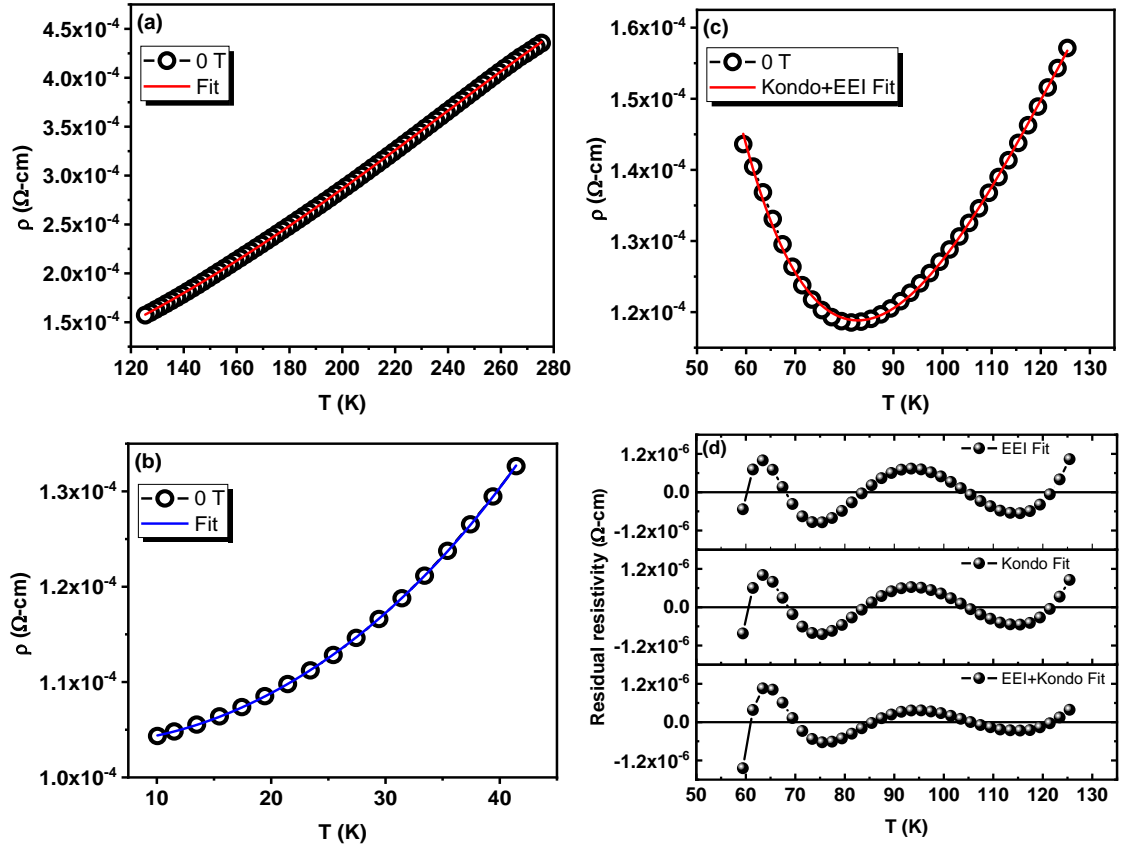

**Figure 8**. The temperature-dependent resistivity fits within the temperature regimes of (a) 125-275 K, (b) 10- 40 K under no magnetic field using Equation (4), (c) 59-125 K under no magnetic field using Equation (7), and (d) the difference between the experimental data and theoretical fits using Equation (4), (6), and (7), respectively at the temperature range of 59 -125 K for MnP-400C.

**Table 1**. The fitting parameter values obtained for different magnetic fields of 0, 2, 3, and 7 T over the temperature range of 125-275 K using Eq. (4).

| $\mu_0 H$ (T) | $\rho_0$ (x $10^{-5}$ $\Omega$ cm) | $\rho_{e-Ph}$ (x $10^{-17}$ $\Omega$ cm/K$^{-1}$) | $\rho_{e-e}^{el}$ (x $10^{-9}$ $\Omega$ cm/K$^{-2}$) | $\rho_{e-e/M}$ (x $10^{-15}$ $\Omega$ cm/K$^{-4.5}$) | $r^2$ |
|---|---|---|---|---|---|
| 0 | 6.1824 ± 0.1005 | 7.3457 ± 1.8176 | 6.3446 ± 0.0799 | -2.3309 ± 0.3526 | 0.999965 |
| 2 | 8.3632 ± 0.1156 | 4.3354 ± 2.1785 | 4.9055 ± 0.0937 | -1.4879 ± 0.4214 | 0.999970 |
| 3 | 10.0288 ± 0.1333 | 6.5664 ± 2.3685 | 4.8484 ± 0.1049 | -1.9242 ± 0.4599 | 0.999966 |
| 7 | 12.4831 ± 0.2368 | 43.6253 ± 4.4974 | 6.6114 ± 0.1929 | -9.4143 ± 0.8696 | 0.999838 |

**Table 2**. The fitting parameter values obtained for different magnetic fields of 0, 2, 3, and 7 T over the temperature range of 10-44 K using Eq. (4).

| $\mu_0 H$ (T) | $\rho_0$ (x $10^{-4}$ $\Omega$ cm) | $\rho_{e-Ph}$ (x $10^{-13}$ $\Omega$ cm/K$^{-1}$) | $\rho_{e-e}^{el}$ (x $10^{-8}$ $\Omega$ cm/K$^{-2}$) | $\rho_{e-e/M}$ (x $10^{-12}$ $\Omega$ cm/K$^{-3.5}$) | $r^2$ |
|---|---|---|---|---|---|
| 0 | 1.0247 ± 0.1691 | 1.6964 ± 0.7944 | 1.6725 ± 0.0877 | -1.0014 ± 0.5766 | 0.999739 |
| 2 | 0.8644 ± 0.0026 | 1.4535 ± 0.9997 | 1.6594 ± 0.1293 | -1.0189 ± 0.7481 | 0.999770 |
| 3 | 0.8476 ± 0.0023 | 1.6383 ± 0.8680 | 1.6906 ± 0.1123 | -1.1811 ± 0.6495 | 0.999821 |
| 7 | 0.7660 ± 0.0012 | 1.6267 ± 0.4987 | 1.7182 ± 0.0615 | -1.1998 ± 0.3703 | 0.999947 |

# Supporting Information

# The Discovery of Giant Positive Magnetoresistance in Proximity to Helimagnetic Order in Manganese Phosphide Nanostructured Films

Nivarthana W.Y.A.Y. Mudiyanselage[1], Derick DeTellem[1], Amit Chanda[1], Anh Tuan Duong[2], Tzung-En Hsieh[3,5], Johannes Frisch[3,5], Marcus Bär[3,4,5,6], Richa Pokharel Madhogaria[7], Shirin Mozaffari[7], Hasitha Suriya Arachchige[7], David Mandrus[7], Hariharan Srikanth[1], Sarath Witanachchi[1], and Manh-Huong Phan[1,*]

[1] Department of Physics, University of South Florida, Tampa, Florida 33620, USA

[2] Phenikaa Research and Technology Institute, Phenikaa University, Yen Nghia, Ha-Dong District, Hanoi, 10000, Vietnam

[3] Department of Interface Design, Helmholtz-Zentrum Berlin für Materialien und Energie GmbH (HZB), 12489 Berlin, Germany

[4] Department of Chemistry and Pharmacy, Friedrich-Alexander-Universität Erlangen-Nürnberg (FAU), 91058 Erlangen, Germany

[5] Energy Materials In-situ Laboratory Berlin (EMIL), HZB, 12489 Berlin, Germany

[6] Department X-ray Spectroscopy at Interfaces of Thin Films, Helmholtz Institute Erlangen-Nürnberg for Renewable Energy (HI ERN), 12489 Berlin, Germany

[7] Department of Materials Science and Engineering, University of Tennessee, Knoxville, Tennessee 37996, USA

*Corresponding author: phanm@usf.edu

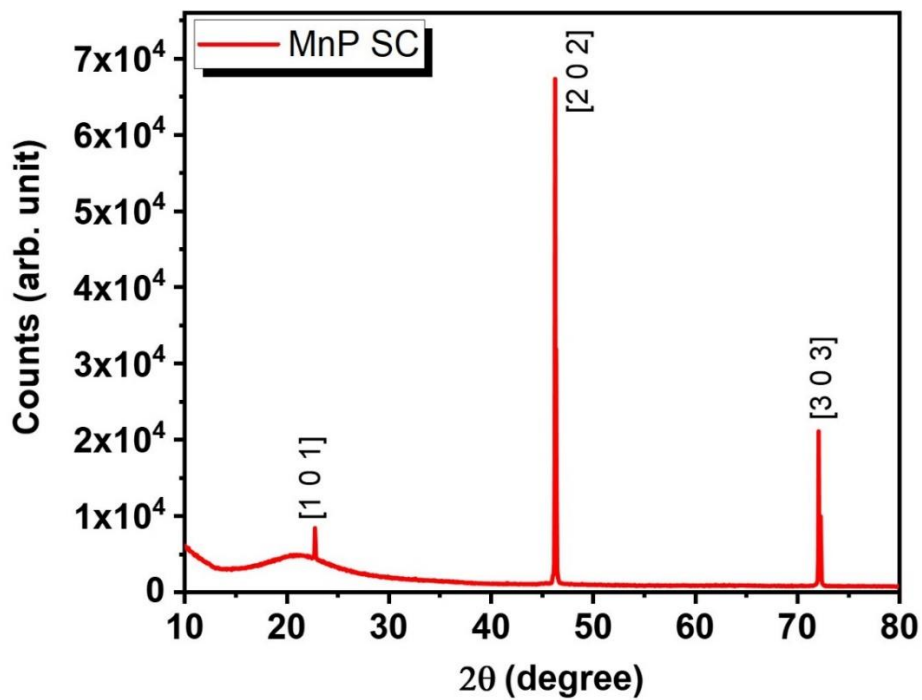

**Figure S1**. The XRD pattern of the MnP single crystal.

The XRD peaks indicate the parallel planes of [1 0 1] planes at 22.790 (identified as [1 0 1]), 46.310 (identified as [2 0 2]) and 72.130 (identified as [3 0 3]). The crystal structure is orthorhombic, and the space group is Pnma with lattice constants of $a$ = 5.242 Å, $b$ = 3.180 Å, and $c$ = 5.903 Å.

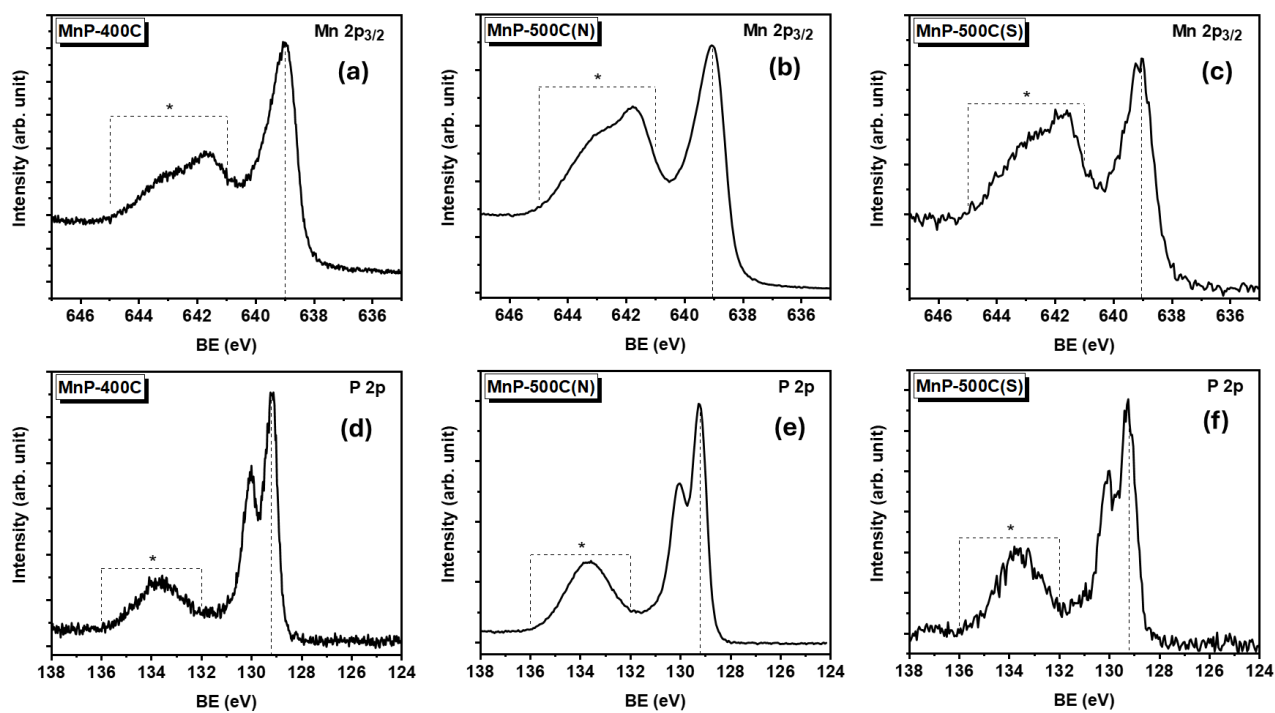

**Figure S2.** Mn $2p_{3/2}$ (a-c) and P 2p (d-f) XPS spectra for the three different MnP samples recorded using a monochromatized Al K$_\alpha$ source (hν=1486.58 eV). The vertical lines at 639.0 eV and 129.2 eV indicate the position of the Mn $2p_{3/2}$ and the P $2p_{3/2}$ peaks, respectively, for Mn-P bonds in MnP [1–3]. The high binding energy contributions (indicated with asterisk brackets) are attributed to the corresponding oxidized species.

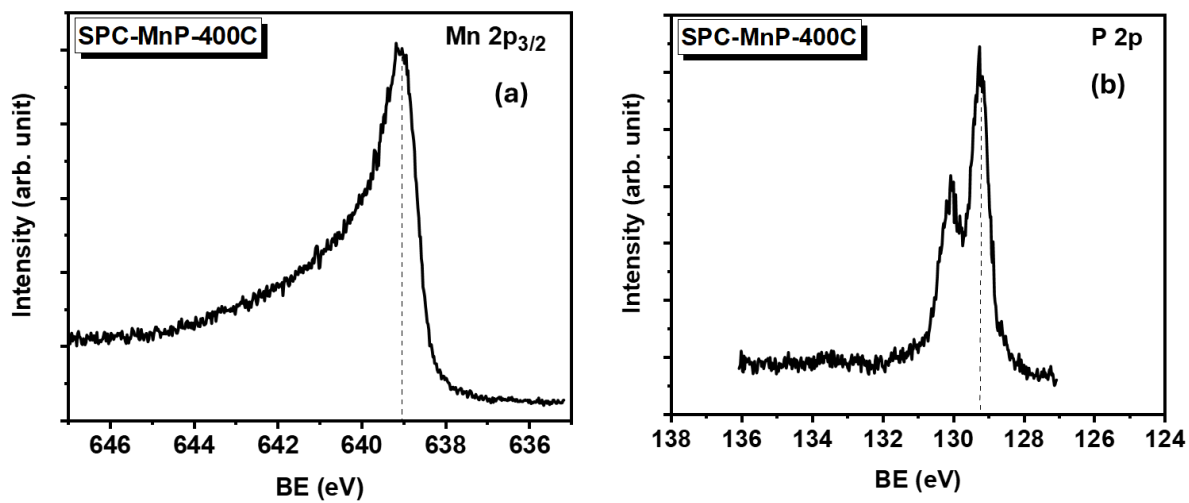

**Figure S3.** Mn 2p$_{3/2}$ (a) and P 2p (b) XPS spectra of the sputter-cleaned (SPC) MnP-400C film recorded using a monochromatized Al K$_\alpha$ source (hν=1486.58 eV). The vertical lines at 639.0 eV and 129.2 eV indicate the position of the Mn 2p$_{3/2}$ and the P 2p$_{3/2}$ peaks, respectively, for Mn-P bonds in MnP [1–3].

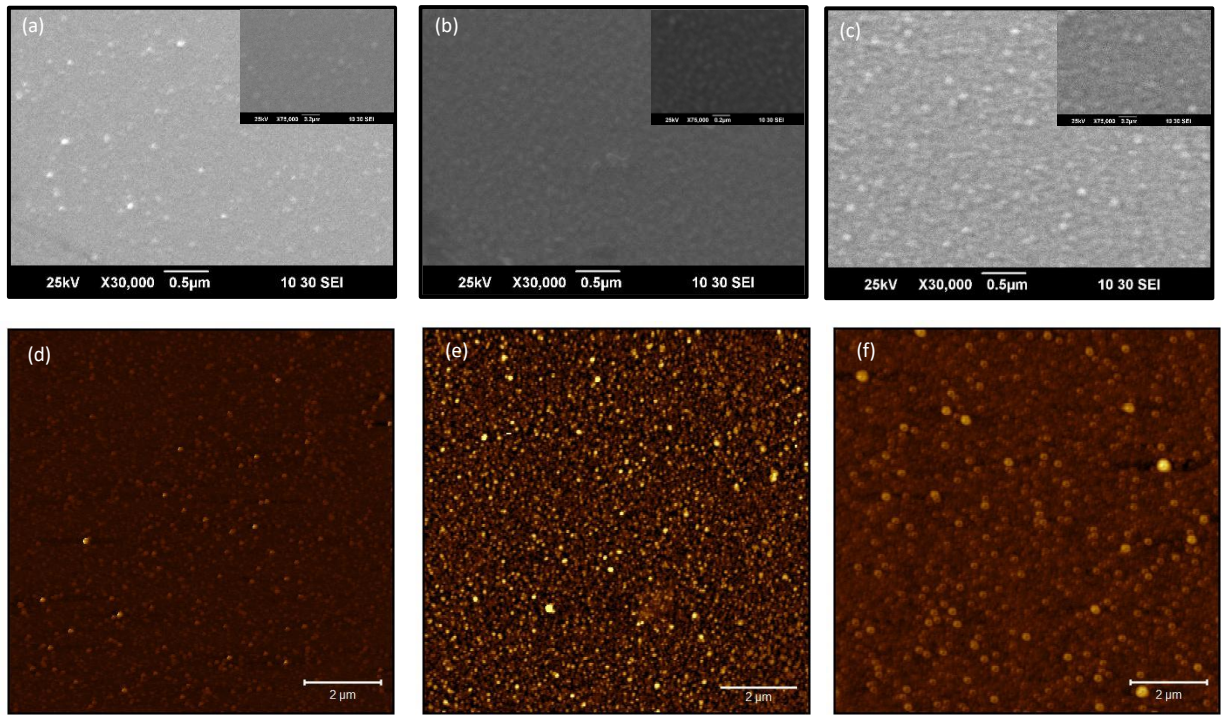

**Figure S4**. SEM images of (a) MnP-400C, (b) MnP-500C(N), and (c) MnP-500C(S); The AFM images of (d) MnP-400C, (e) MnP-500C(N), and (f) MnP-500C(S).

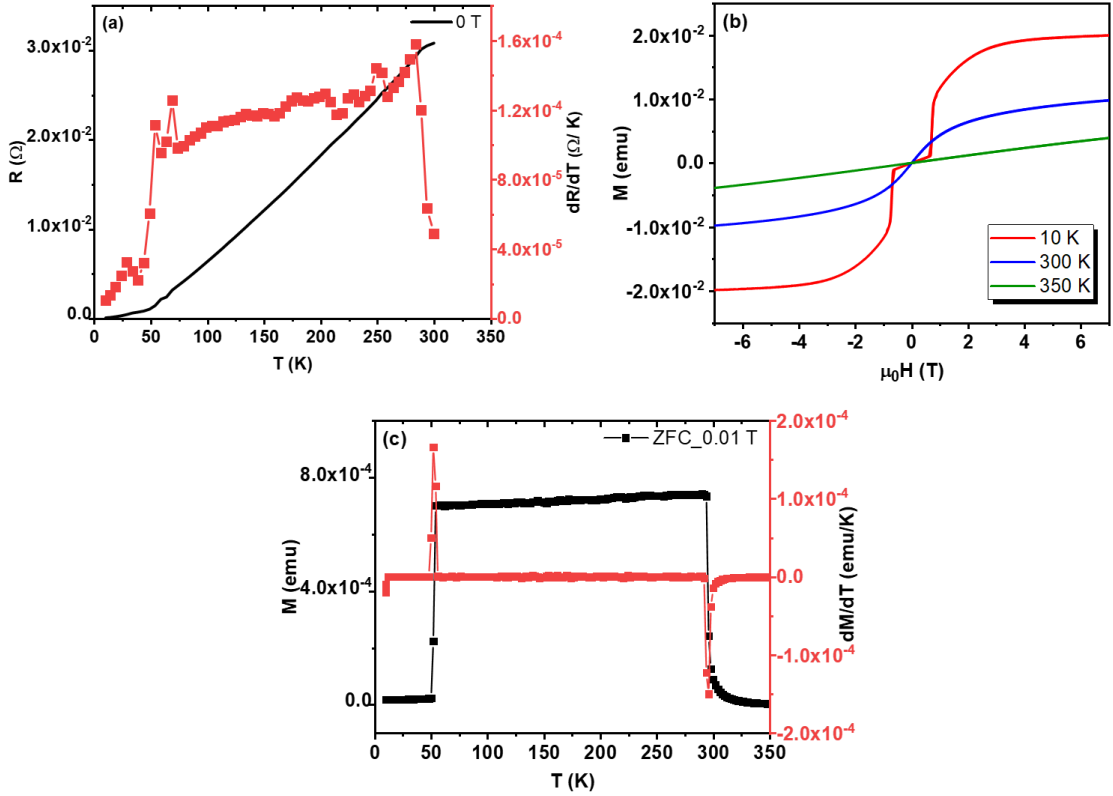

**Figure S5**. (a) Temperature-dependent resistance and its derivative of the MnP single crystal; (b) Magnetic field-dependent magnetization $M(H)$ curves measured at different temperatures for the same sample; (c) Temperature-dependent magnetization $M(T)$ under ZFC protocol.

Note that the MnP single crystal exhibits a paramagnetic to ferromagnetic (PM-FM) phase transition at ~296 K followed by another ferromagnetic to helical (FM-HM) phase transition at ~52 K. The temperature-dependent resistivity exhibits a kink around the PM-FM transition temperature, which is further manifested as a jump in the derivative of resistivity with respect to temperature. Similarly, there is another kink around at ~53 K, which is related to the FM-HM transition.

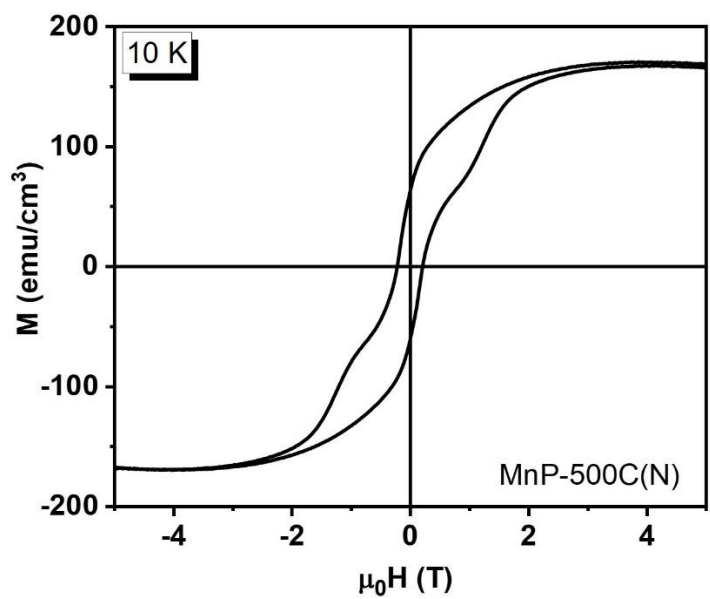

**Figure S6.** Isothermal in-plane magnetic hysteresis $M(H)$ loop at 10 K for MnP-500C(N).

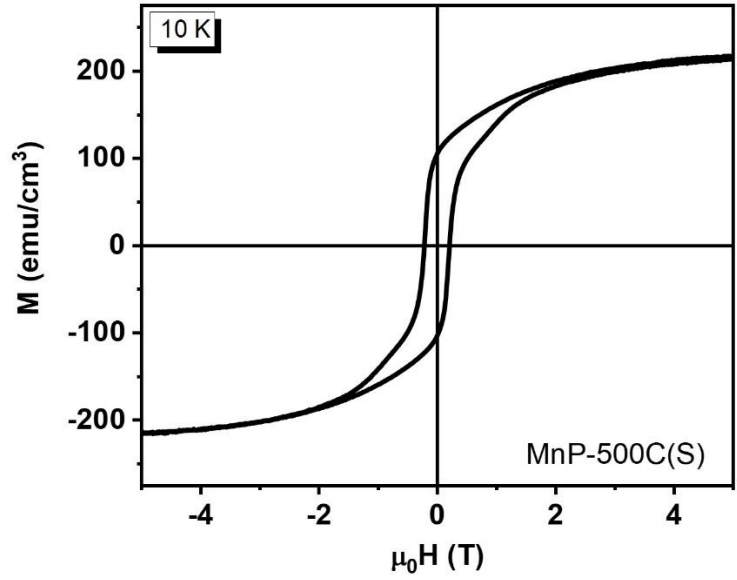

**Figure S7.** Isothermal in-plane magnetic hysteresis *M*(*H*) loop measured at 10 K for MnP-500C(S).

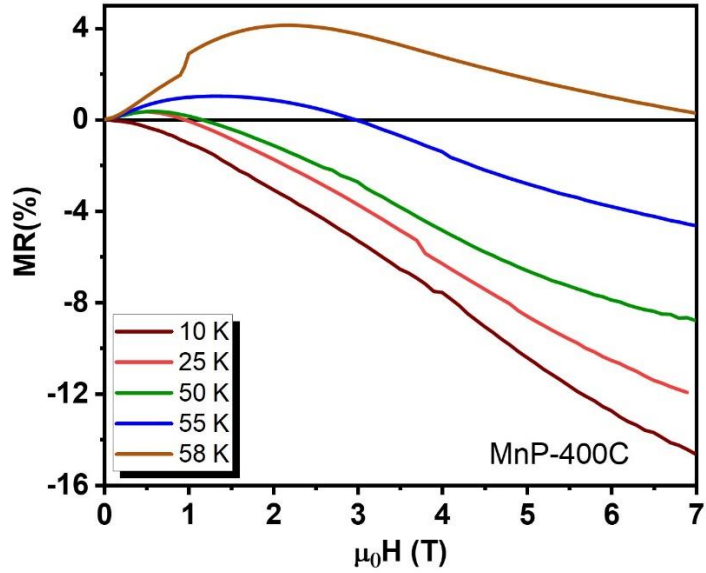

**Figure S8.** Magnetic field-dependent magnetoresistance (MR) ratio (%) of MnP-400C.

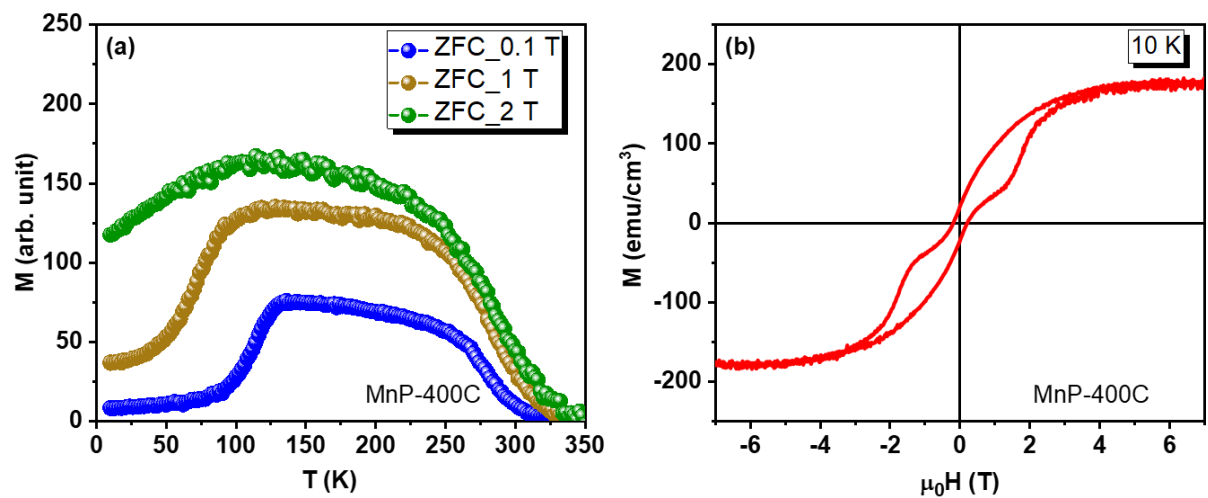

**Figure S9.** (a) Temperature-dependent magnetization $M(T)$ of MnP-400C under zero-field-cooled measurement protocol for different applied fields; (b) Isothermal in-plane magnetic hysteresis $M(H)$ loop at 10 K for MnP-400C.

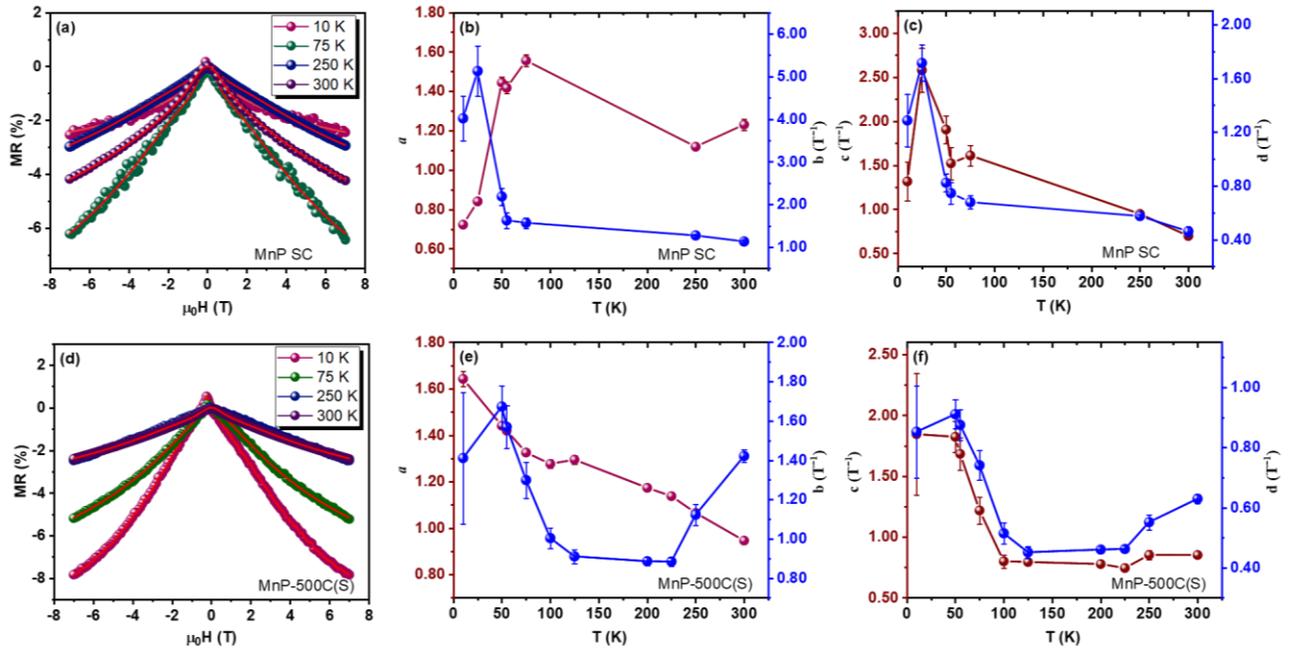

**Figure S10.** (a), (d) The magnetic field-dependent MR ratio (%) at representative temperatures, 10 K, 75 K, 250 K, and 300 K for the MnP single crystal and MnP-500C(S), respectively. The solid (red) lines indicate the fitting curves using Equation (2) in the main text; (b), (e) *a* and *b* parameters and (c), (f) *c* and *d* parameters obtained by fitting the experimental MR data using Equation (2) for MnP SC and MnP-500C(S).

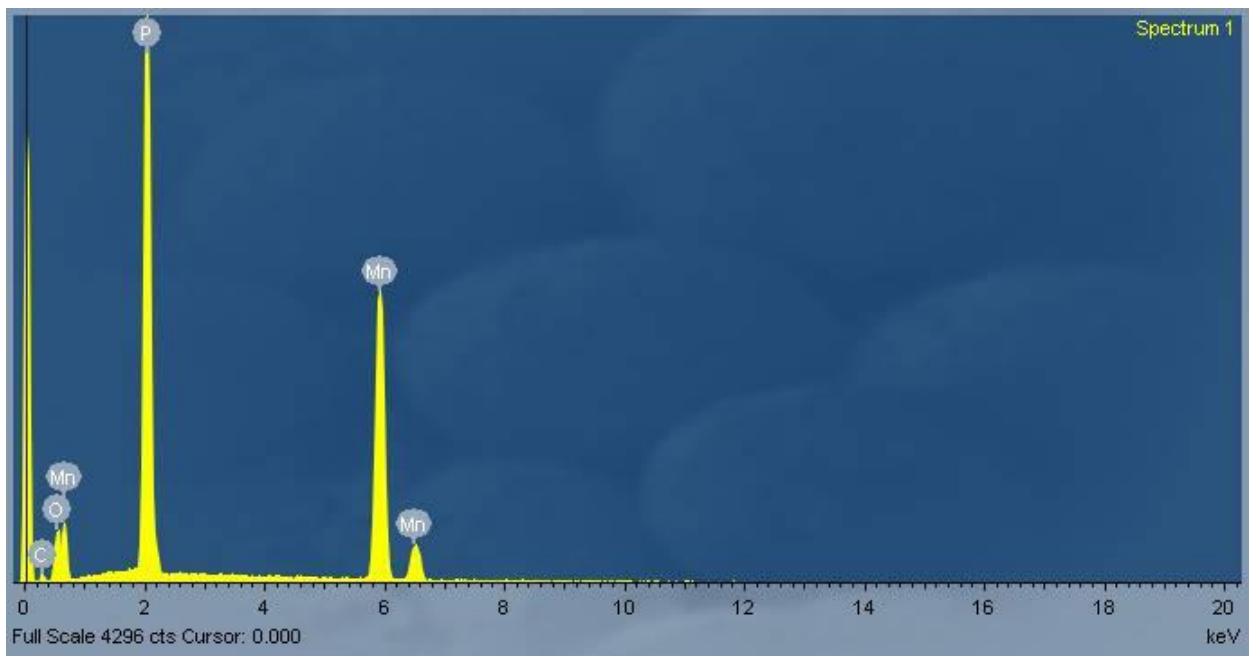

**Figure S11.** Energy dispersive spectroscopy (EDS) of the MnP single crystal confirmed the presence of Mn and P. The O and C signals are attributed to surface oxidation and/or contamination resulting from exposure to air.

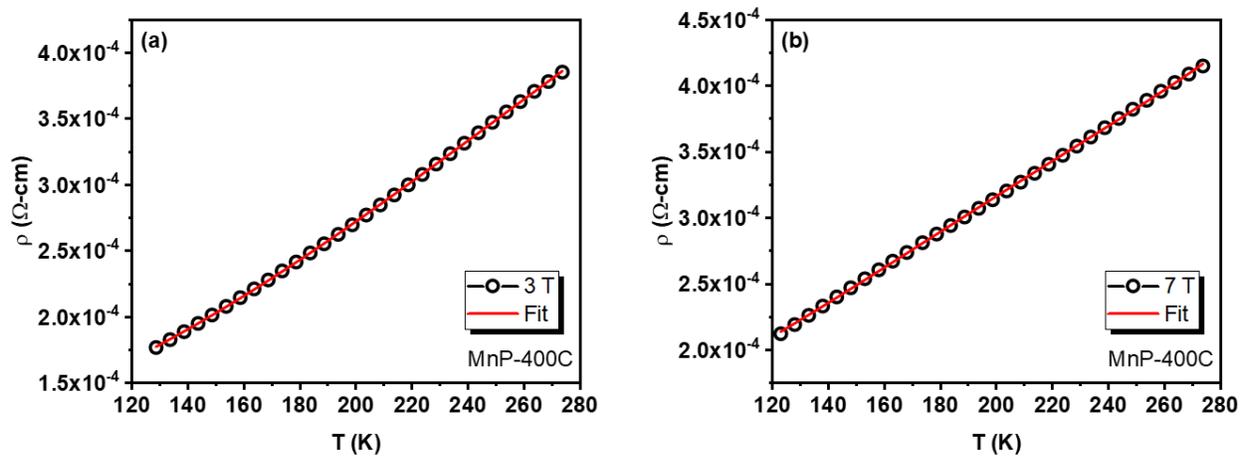

**Figure S12.** Resistivity fits for the temperature range of 125-275 K for MnP-400C under magnetic fields of (a) 3 T and (b) 7 T using Equation (4) given in the main text.

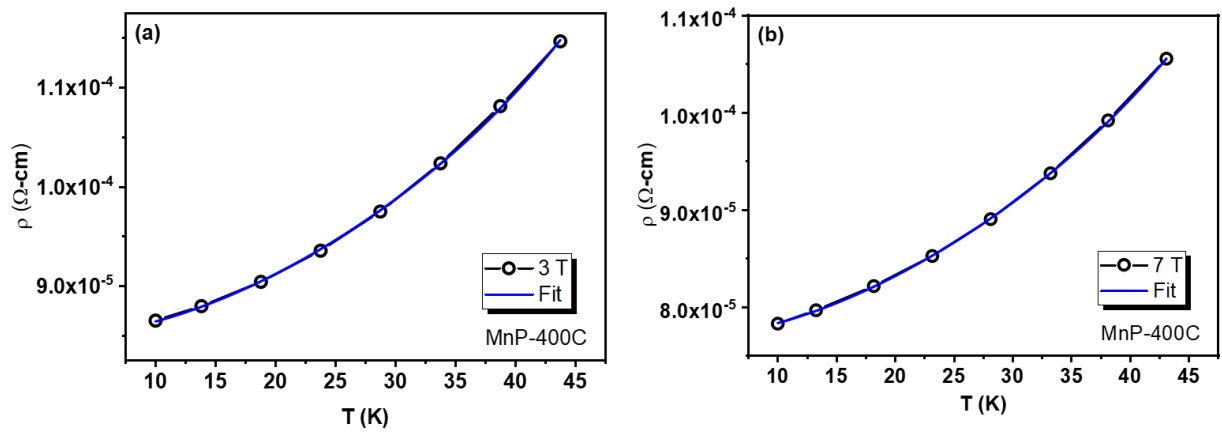

**Figure S13.** Resistivity fits for the temperature range of 10-44 K for MnP-400C under magnetic fields of (a) 3 T and (b) 7 T using Equation (4) given in the main text.

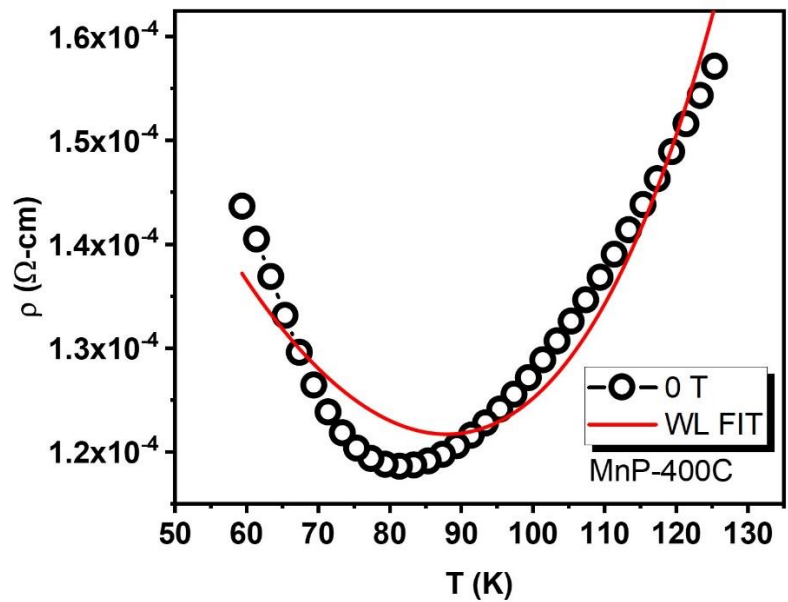

**Figure S14.** Temperature-dependent resistivity under zero magnetic field for the 59 K-125 K region for MnP-400C. The solid line represents the fit using Equation (5) given in the main text.

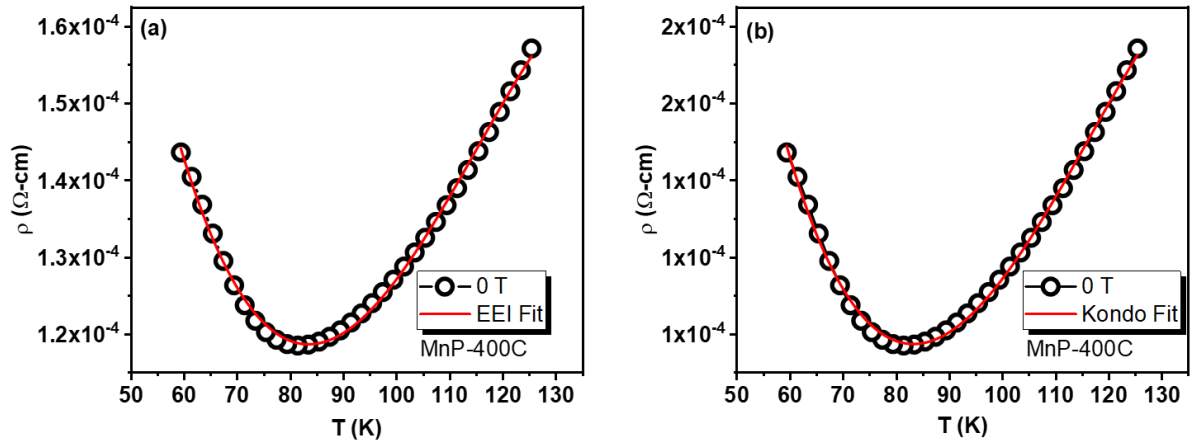

**Figure S15.** The temperature-dependent resistivity fits within the temperature regime of 59-125 K under no magnetic field (a) using Equation (4) and (b) using Equation (6) given in the main text for MnP-400C.

**Table S1.** Magnetic ordering temperatures of all MnP samples

| Sample | $T_N$ (K) | $T_C$ (K) |
|---|---|---|
| MnP SC | ~ 52 | ~ 296 |
| MnP-500C(S) | ~ 74 | ~ 304 |
| MnP-500C(N) | ~ 102 | ~ 300 |
| MnP-400C | ~ 112 | ~ 278 |

**Table S2.** The fitting parameters obtained using Equation (4), (6) and (7) for MnP-400C.

| Fit | $\rho_0$ (x $10^{-4}$ $\Omega$ cm) | $\rho_{e-Ph}$ (x $10^{-15}$ $\Omega$ cm/K$^{-5}$) | $\rho^{el}_{e-e}$ (x $10^{-4}$ $\Omega$ cm/K$^{-2}$) | $\rho_{e-e/M}$ (x $10^{-8}$ $\Omega$ cm/K$^{-2}$) | $\rho_K$ (x $10^{-3}$ $\Omega$ cm/K$^{-2}$) | $r^2$ | Chi-square (x $10^{-13}$) |
|---|---|---|---|---|---|---|---|
| EEI Fit | 8.0024 ± 0.1709 | -3.1895 ± 0.1865 | -1.0229 ± 0.0270 | 3.8169 ± 0.1186 | N/A | 0.9973 | 3.9951 |
| Kondo + EEI Fit | 27.1000 ± 0.5992 | -67.4124 ± 0.0868 | 2.4061 ± 0.0627 | N/A | -1.0800 ± 0.0264 | 0.9983 | 2.4950 |
| Kondo | 13.7000 ± 0.2900 | -2.4480 ± 0.1513 | N/A | 2.68319 ± 0.0802 | -32.3041 ± 0.0077 | 0.9978 | 3.2616 |

**Table S3.** The fitting parameters obtained using Equation (2) for MnP-500C(N).

| T (K) | a | b (T$^{-1}$) | c (T$^{-1}$) | d (T$^{-1}$) |
|---|---|---|---|---|
| 50 | 1.6519 ± 0.0082 | 1.6658 ± 0.0842 | 2.0843 ± 0.1125 | 0.8956 ± 0.0349 |
| 55 | 1.5932 ± 0.0070 | 1.74607 ± 0.0785 | 2.1139 ± 0.0985 | 0.9224 ± 0.0313 |
| 75 | 0.9933 ± 0.0066 | 2.0239 ± 0.0880 | 1.5494 ± 0.0511 | 0.8940 ± 0.0251 |